\documentclass[aps,prx,twocolumn,showpacs,superscriptaddress]{revtex4-2}  
\usepackage{graphicx}  % needed for figures
\usepackage{dcolumn}   % needed for some tables
\usepackage{bm}        % for math
\usepackage{amssymb}   % for math
\usepackage{amsmath}   % for math

\usepackage{verbatim}  % comments 
\usepackage{xcolor}    % color of text

\begin{document}
\title{Mass-induced Coulomb drag in capacitively coupled superconducting nanowires}

\author{Aleksandr Latyshev}
\affiliation{Laboratoire de Physique des Solides, CNRS UMR 5802-University Paris-Saclay, France}
\affiliation{D\'epartement de Physique Th\'eorique, Universit\'e de Gen\`eve, CH-1211 Gen\`eve 4, Switzerland}

\author{Adrien Tomà}
\affiliation{D\'epartement de Physique Th\'eorique, Universit\'e de Gen\`eve, CH-1211 Gen\`eve 4, Switzerland}

\author{Eugene V. Sukhorukov}
\affiliation{D\'epartement de Physique Th\'eorique, Universit\'e de Gen\`eve, CH-1211 Gen\`eve 4, Switzerland}
\date{\today}

\begin{abstract}
We investigate Coulomb drag in a system of two capacitively coupled superconducting nanowires. In this context, drag refers to the appearance of a stationary voltage in the passive wire in response to a current bias applied to the active one. Quantum phase slips (QPS) in the biased wire generate voltage fluctuations that can be transmitted to the other. Using perturbative and semiclassical approaches, we show that when both wires are superconducting the induced voltage vanishes due to exact cancellation of plasmon contributions. By contrast, when the second wire is tuned below the superconductor–insulator transition and develops a mass gap, this cancellation is lifted and a finite drag voltage emerges. The drag coefficient exhibits a crossover from weak drag in short wires to a maximal value set by the mutual capacitance in long wires. A semiclassical picture of voltage pulse propagation clarifies the physical origin of the effect: the mass term synchronizes plasmon modes and prevents complete cancellation of induced signals. Our results establish a mechanism of mass-induced Coulomb drag in low-dimensional superconductors and suggest new routes for probing nonlocal transport near quantum criticality.
\end{abstract}

\maketitle

\section{Introduction}

Fluctuations play a decisive role in low-dimensional superconductors, where reduced dimensionality enhances the impact of collective excitations and quantum effects. Superconductivity is conventionally understood as the spontaneous breaking of the global $U(1)$ symmetry, giving rise to collective amplitude and phase modes of the order parameter $\Delta(x,t) = \Delta_0(x,t)e^{i\phi(x,t)}$. In bulk systems, phase coherence is robust, but in narrow superconducting wires with transverse dimensions below the coherence length $\xi$, fluctuations of the phase $\phi$ dominate at low temperatures. In particular, quantum phase slips (QPS) — topological events in which the superconducting order parameter temporarily vanishes and the phase jumps by $\pm 2\pi$ — provide a key mechanism for the breakdown of superconductivity in one dimension \cite{AGZ08,bezryadin08,zaikin10,bezryadin2013,book}. Each QPS generates voltage pulses via the Josephson relation, linking microscopic fluctuations to measurable transport signatures.

The physics of QPS can be recast in terms of a two-dimensional Coulomb gas of vortices in Euclidean space-time \cite{ZGO}. This mapping highlights the logarithmic interactions between QPS events and the emergence of a Berezinskii–Kosterlitz–Thouless (BKT)–type transition \cite{berezinskii71,kosterlitz18,kosterlitz74}, in which the proliferation of unbound phase slips drives the destruction of long-range order. Experimental advances in the fabrication of ultrathin superconducting nanowires \cite{BLT2000,lau2001,baumans2016,arutyunov2021} have confirmed this scenario, demonstrating the suppression of superconductivity and the onset of insulating behavior at low temperatures. These developments have spurred a broad interest in QPS as a fundamental ingredient of one-dimensional superconductivity and as a resource for novel quantum devices.

A complementary theoretical description is provided by the Sine–Gordon model  \cite{tsvelik_book,diehl1997sine,semenov2013}, which captures the tunneling of flux quanta across the wire and allows for a controlled perturbative treatment in the superconducting regime. Within this framework, QPS not only generate a finite average voltage but also produce non-equilibrium shot noise \cite{SZ2016}. Such transport signatures provide a natural link between microscopic phase fluctuations and macroscopic observables. Extensions of this approach to coupled systems have revealed additional collective phenomena. In particular, studies of capacitively coupled nanowires \cite{LSZ20} have shown that QPS in one wire induce correlated voltage fluctuations in the other. While these fluctuations manifest as stochastic voltage pulses, the average induced voltage vanishes due to the dynamical screening provided by collective plasma modes \cite{LSZ}.

The central question addressed in this work is how this picture is modified when one of the coupled wires is tuned through the superconductor–insulator transition (SIT) and develops a finite mass gap. In this regime, QPS proliferate in the insulating wire, and its low-energy Hamiltonian acquires a quadratic “massive” term for the plasmon field. We demonstrate that this modification has profound consequences for nonlocal transport: it prevents the complete cancellation of induced voltage pulses, thereby producing a finite Coulomb drag effect mediated by QPS in the superconducting wire. This drag mechanism is absent in the purely superconducting regime and emerges as a direct consequence of the mass gap.

To establish this result, we employ a combination of perturbative and semiclassical approaches. The perturbative analysis treats QPS in the superconducting wire as a weak process, while the semiclassical description provides an intuitive picture of voltage pulse dynamics in the presence of a mass gap. Together, these methods demonstrate how dimensionality, mutual capacitance, and the opening of a plasmonic gap work together to produce a robust drag response. Beyond its fundamental interest, the phenomenon of mass-induced Coulomb drag opens a route to probing correlated dynamics in superconducting nanostructures and offers new perspectives on nonlocal transport in low-dimensional systems.

%=================================================================================
\section{Model of the two capacitively coupled superconducting nanowires}

\begin{figure}[h]
\centering
\includegraphics[width=0.5\textwidth]{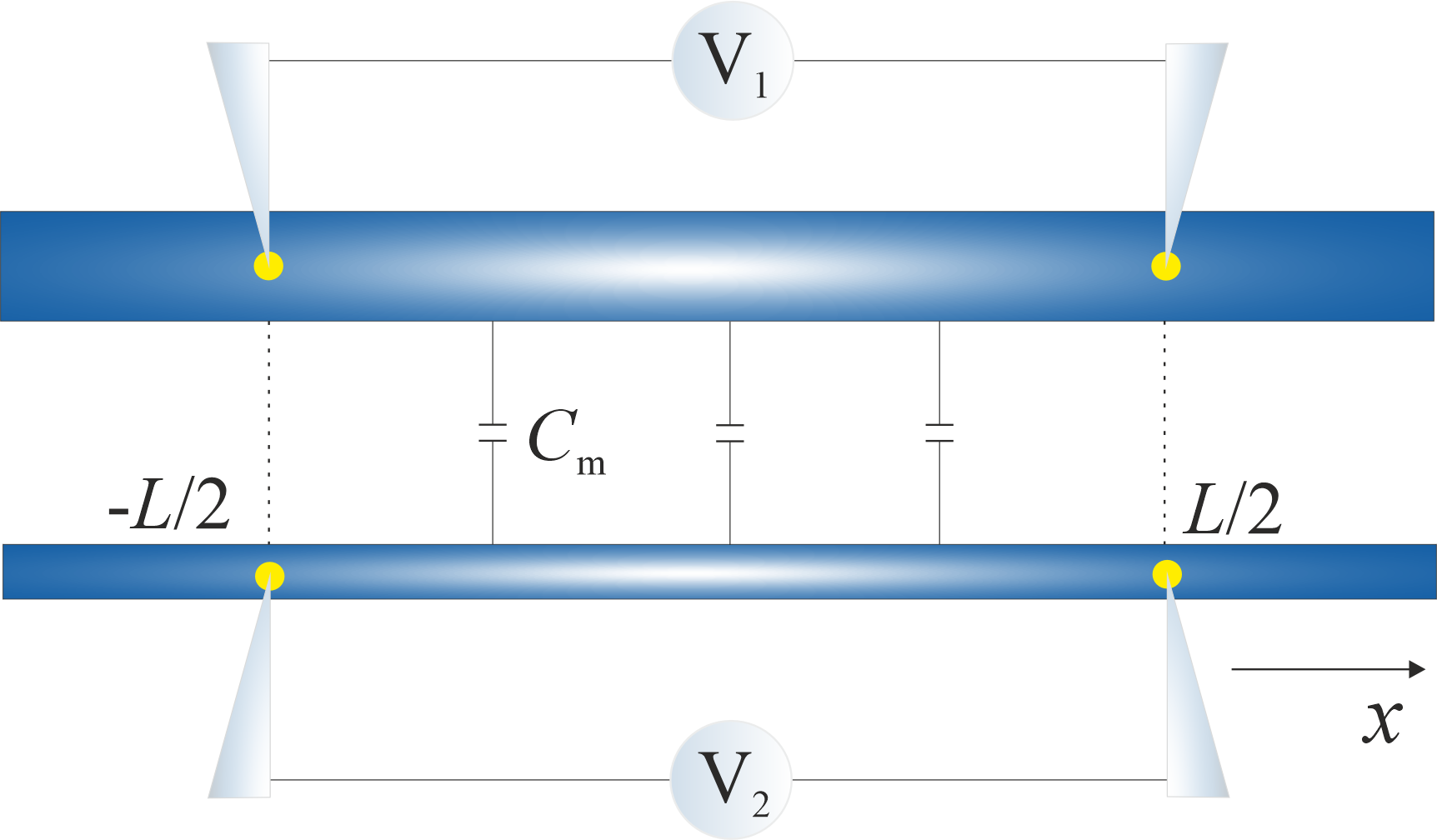}
\caption{Schematic representation of the system of two capacitively coupled superconducting
nanowires. The notations are explained in the main text.}
\label{two wires}
\end{figure}

We consider two sufficiently thin superconducting nanowires that are capacitively coupled, as illustrated schematically in Fig.\ \ref{two wires}.
The wires are characterized by a geometric capacitance $C_{1(2)}$ (per unit length) and kinetic inductance $\mathcal{L}_{1(2)}$ (multiplied by  unit length), effectively forming two transmission lines with the Hamiltonian
\begin{multline}\label{TL1}
\hat{H}_{TL} = \frac{1}{2} \sum_{i,j=1,2} \int dx \Bigg( \mathcal{L}^{-1}_{ij} \hat{\Phi}_{i}(x) \hat{\Phi}_{j}(x) \\ +C^{-1}_{ij} \frac{\nabla \hat{\chi}_{i}(x) \nabla \hat{\chi}_{j}(x)}{\Phi_{0}^{2}} \Bigg),
\end{multline}
where $\Phi_{0}=\pi/e$ is the flux quantum,  and $C_{ij}$ and $\mathcal{L}_{ij}$ are the matrix elements of the  capacitance and inductance matrices,
\begin{eqnarray}
    \check{C} =
    \begin{pmatrix}
    C_1 & -C_m \\
    -C_m & C_2
    \end{pmatrix}, \;\;\;\;
    \check{\mathcal{L}} = 
    \begin{pmatrix}
    \mathcal{L}_1 & 0 \\
    0 & \mathcal{L}_2
    \end{pmatrix}.
\end{eqnarray}
The Coulomb interaction between the two wires is represented by the off-diagonal (mutual) capacitance $C_{m}$ in $\check{C}$.

The Hamiltonian (\ref{TL1}) is expressed in terms of the canonically conjugate operators $\hat{\Phi}$ and $\hat{\chi}$ with the commutation relation
\begin{eqnarray}\label{comm}
    [\hat{\Phi}_i(x), \hat{\chi}_j(x')] = -i \Phi_0 \delta_{ij} \delta(x - x'),
\end{eqnarray}
which are related to the phase $\hat{\varphi}$ and the charge density $\hat{\rho}$ operators through
\begin{eqnarray}\label{duality}
 \hat{\rho}_{i}(x) = \frac{1}{\Phi_{0}} \partial_{x} \hat{\chi}_{i}(x),\;\; \hat{\varphi}_{i}(x)=2e \int^{x}_{0} dy \hat{\Phi}_{i}(y),    
\end{eqnarray}
obeying the canonical commutation relations
\begin{eqnarray}\label{CR}
 [\hat{\rho}_{i}(x),\hat{\varphi}_{j}(x^{\prime})]=-2e \delta_{ij}\delta(x-x^{\prime}).   
\end{eqnarray}

If the wires are sufficiently thick, strong phase fluctuations (phase slips) can be neglected, and the low-energy Hamiltonian (\ref{TL1}) is sufficient to describe the dynamics of collective excitations. For thinner wires, however, QPS must also be taken into account. Their contribution to the Hamiltonian can be written as \cite{semenov2013,book}
\begin{eqnarray}
 \hat{H}_{QPS}&=&\hat{H}_{QPS_{1}}+\hat{H}_{QPS_{2}}, \\
 \hat{H}_{QPS_{i}}&=&- \gamma_i\int dx\cos(\hat{\chi}_{i}(x)),\;\; i=1,2,
 \label{QPS}
\end{eqnarray}
where the coupling constants
\begin{eqnarray}\label{amplitude}
  \gamma_{i} \sim  (g_{i\xi}\Delta/\xi)\exp(-a g_{i\xi})
\end{eqnarray}
represent the QPS amplitudes per unit length. Here $\Delta$ is the superconducting order parameter, $g_{j\xi} = R_q / R_{j\xi}$ is the dimensionless conductance, $R_{j\xi}$ is the normal-state resistance of a wire segment of length $\xi$, and $a$ is a numerical constant of order one. Since $g_{i\xi}\sim \sqrt{s_{i}}$ depends on the cross-sectional area $s_{i}$ of the wire, the QPS contribution is exponentially suppressed for wires with $\sqrt{s_{i}} \sim \xi$.  The total Hamiltonian of the system is therefore
\begin{eqnarray}
\hat{H}=\hat{H}_{TL}+\hat{H}_{QPS_{1}}+\hat{H}_{QPS_{2}}. 
\end{eqnarray}

We first assume that the first wire is relatively thick (but still $\sqrt{s_{1}} < \xi$), so that the dimensionless parameter characterizing QPS interactions satisfies $\lambda_{1}=\pi /(4e^{2})\sqrt{C_{1}/\mathcal{L}_{1}} > 2$. In this case, QPS and anti-QPS form bound pairs, and the first wire remains globally superconducting. The amplitude $\gamma_{1}$ in (\ref{QPS}) is therefore small and can be treated perturbatively. In contrast, the second wire is assumed to be sufficiently thin such that $\lambda_{2} < 2$. Here, single QPS events dominate, strongly suppressing superconductivity. In this regime, one may employ the semiclassical limit \cite{IC10,foini2015,foini2017,ruggiero2021}, where the potential (\ref{QPS}) is expanded around one of its minima \cite{IC10}, so that the cosine interaction can be approximated by a harmonic potential in $\hat{\chi}_2$. This semiclassical description can also be justified by a central limit theorem argument: with many QPS events, their distribution tends toward a Gaussian. The resulting Hamiltonian for the second wire takes the quadratic form
\begin{eqnarray}
\hat{H}_{QPS_{2}}=\frac{\gamma_{2}}{2}\int dx [\hat{\chi}_{2}(x)]^{2}.  
\label{mass}
\end{eqnarray}
Therefore, the combined Hamiltonian $\hat{H}_{TL}+\hat{H}_{QPS_{2}}$ is Gaussian, while $\hat{H}_{QPS_{1}}$ remains nonlinear and will be treated perturbatively.

Physically, this quadratic contribution corresponds to a finite plasmonic mass in the second wire, $m^{2}=\gamma_{2}\Phi_{0}^{2}$, reflecting the proliferation of QPS on the insulating side of the superconductor–insulator transition. Earlier analyses of Coulomb drag between superconducting wires did not include such a mass term. As we demonstrate below, the presence of this mass crucially modifies the long-wavelength dynamics and the character of screening, thereby enabling a finite drag response.

The central observable in our analysis is the potential drop $V_i$ between the cross-sections of the two wires at $x=\pm L/2$, as indicated in Fig.\ \ref{two wires}. Varying the Hamiltonian (\ref{TL1}) with respect to the density field (\ref{duality}) yields
\begin{equation}\label{voltage}
\hat{V}_i(t)=\frac{1}{\Phi_{0}}\sum_{j=1,2}C^{-1}_{ij}[\partial_x \hat{\chi}_j(L/2,t) - \partial_x \hat{\chi}_j(-L/2,t)],
\end{equation}
while the current operator follows from the continuity equation,
\begin{equation}
\hat{I}_i(x,t) =\frac{1}{\Phi_0}\partial_t \hat{\chi}_i(x,t).
\end{equation}
A current bias $I_1$ applied to the first nanowire can be incorporated as a simple shift,
\begin{eqnarray}\label{shift}
    \hat{\chi}_{1}(x,t) \to I_{1}\Phi_{0}t+\hat{\chi}_{1}(x,t),
\end{eqnarray}
which leaves the Gaussian part of the Hamiltonian unchanged but modifies the vertex operators in Eq.\ (\ref{QPS}).

%=================================================================================
\section{Perturbation theory}
\label{Spectral density of current noise}

Since the QPS contribution in the first wire is irrelevant, we can analyze its effect perturbatively and study how QPS processes influence voltage fluctuations in both wires. For the second wire, however, perturbation theory is not applicable because the QPS contribution is relevant. In this case, QPS generate a mass term for the field $\hat{\chi}_{2}$, which must be treated non-perturbatively.

For such a system, it is convenient to use operator perturbation theory, where the average voltage takes the form
\begin{multline}\label{volt1}
  \langle \hat{V}_{i}(t) \rangle=\text{Tr}\Big(\hat{\rho}(-\infty)\tilde{\mathcal{T}}e^{i\int^{t}_{-\infty}dt^{\prime}\hat{H}_{QPS_{1}}(t^{\prime})} \\ \times \hat{V}^{0}_{i}(t)\mathcal{T}e^{-i\int^{t}_{-\infty}dt^{\prime}\hat{H}_{QPS_{1}}(t^{\prime})}\Big),
\end{multline}
where $\hat{\rho}(-\infty)$ is the initial density matrix of the system, and $\mathcal{T}$ ($\tilde{\mathcal{T}}$) denotes the time (anti-time) ordering operator. Expression (\ref{volt1}) is written in the interaction representation, with
\begin{equation}
\hat{V}^{0}_{i}(t)=e^{i\hat{H}_{TL}t+i\hat{H}_{QPS_{2}}t}\hat{V}^{0}_{i}(0)e^{-i\hat{H}_{TL}t-i\hat{H}_{QPS_{2}}t}.
\end{equation}
Expanding in  $\hat{H}_{QPS_{1}}$, Eq.\ (\ref{volt1}) reduces at leading non-vanishing order in the QPS amplitude $\gamma_{1}$ \cite{SZ2016,LSZ20} to
\begin{multline}\label{PT}
 \langle \hat{V}_{i}(t) \rangle \approx (-i)^{2} \int^{t}_{-\infty}dt_{1} \int^{t_{1}}_{-\infty} dt_{2} \Big\langle \Big[[\hat{V}^{0}_{i}(t),\hat{H}_{QPS_{1}}(t_{1})],\\
 \times\hat{H}_{QPS_{1}}(t_{2})\Big] \Big\rangle.
\end{multline}
The first-order contribution vanishes because the energy of a single QPS is large (of order of the system size), rendering this contribution exponentially suppressed.

Expanding the commutators in Eq.\ (\ref{PT}), one arrives at the expression for the average voltages obtained in Ref.\ \cite{LSZ20},
\begin{multline}\label{AV}
 \langle\hat{V}_{i} \rangle=\lim_{\omega \to 0} \int^{L/2}_{-L/2} dx \; G_{V_i\chi_1}^{R}(\omega,x) \\ \times\left[\Gamma_{QPS}(I_{1}\Phi_{0})-\Gamma_{QPS}(-I_{1}\Phi_{0})\right],
\end{multline}
where the retarded functions
\begin{eqnarray}\label{RF}
    G_{V_{i}\chi_{j}}^{R}(x,t,t^{\prime})=-i\theta(t-t^{\prime})\langle [\hat{V}_{i}(t),\hat{\chi}_{j}(x,t^{\prime})] \rangle
\end{eqnarray}
describe the response to voltage fluctuations in both wires. The leading-order decay rate $\Gamma_{QPS}(\omega)$ of the current state is derived in Appendix~\ref{Keldysh} [see Eq.\ (\ref{decay})].

In Ref.~\cite{LSZ20}, it was shown that when QPS in the second wire are neglected ($\gamma_{2}=0$), the induced voltage $\langle \hat{V}_{2} \rangle$ vanishes. In our formulation, this corresponds to the limit $m^{2}\to 0$. Mathematically, the vanishing follows from the zero-frequency behavior of the response function \eqref{RF} in Eq.~\eqref{AV}, namely $G^{R}_{V_{2}\chi_{1}}(x,\omega\to 0)\to 0$. Physically, this reflects perfect screening of the charge induced by QPS in the first wire by the second wire. When the second wire develops a gap in its collective mode spectrum, this screening mechanism is qualitatively modified. Although Eq.~(\ref{AV}) has the same formal structure as in the massless case, the plasmonic mass $m$ derived in Sec.~II changes the zero-frequency behavior of the Green functions entering this equation [see, Eq.\  \eqref{EoM}], thereby removing the cancellation mechanism that eliminates drag when both wires are superconducting. The finite induced voltage obtained below therefore follows directly from the mass generated by proliferating QPS in the insulating wire.

Details of the subsequent calculations are presented in Appendix~\ref{Keldysh}. The final result for the voltage in the first wire is
\begin{equation}\label{voltage drop}
    \langle \hat{V}_{1}(I_{1})\rangle = \Phi_{0}L \left[\Gamma_{QPS}(I_{1}\Phi_{0})-\Gamma_{QPS}(-I_{1}\Phi_{0})\right],
\end{equation}
and for the dimensionless drag coefficient, defined as ${\cal D}\equiv  \langle \hat{V}_{2}(I_{1})\rangle/\langle \hat{V}_{1}(I_{1})\rangle$,
\begin{equation}\label{drag}
  {\cal D}=\frac{C_{m}}{C_{2}}\left[1-\frac{1-e^{-L/L_0}}{L/L_0}\right],\quad L_0=1/\sqrt{C_{2}m^{2}},
\end{equation}
where $L_0$ is the characteristic screening length in the second wire. Our theory does not rely on weak interactions. In particular, the prefactor in Eq.\ (\ref{drag}) reaches the maximum value $C_{m}/C_{2}\approx 1$ in the case of long-range Coulomb interactions. The function in brackets describes the crossover from weak drag in short wires, ${\cal D}=(C_{m}/C_{2})\times (L/2L_0)$, to strong drag ${\cal D}=C_{m}/C_{2}$ in long wires. 

Finally, we note that the voltage in the first wire, Eq.\ (\ref{voltage drop}), generally depends on the mass $m$. However, when the applied current or temperature is large enough, $I_{1}\Phi_{0}, T \gg m/\sqrt{\mathcal{L}_{1(2)}}$, the integral \eqref{decay} defining the decay rate $\Gamma_{QPS}$ is dominated by short times $t \ll \sqrt{\mathcal{L}_{1(2)}}/m$. In this regime, the mass term in the retarded and Keldysh Green functions can be neglected. The voltage in the first wire then reduces to
\begin{multline}\label{VVSI}
 \langle \hat{V}_{1}(I_{1})\rangle= \frac{1}{4}\Phi_{0}  u_{\rm eff}\gamma^{2}_{1} \omega_c^2L  \\
 \times\sinh\left(\frac{I_1\Phi_0}{2T}\right)\left(\frac{2\pi T}{\omega_c}\right)^{2\lambda_{11}-2}\\
 \times\frac{\left|\Gamma\left(\frac{\lambda_{11}}{2}+i\frac{I_1\Phi_0}{4\pi T}\right)\right|^4}{\Gamma(\lambda_{11})^2},
\end{multline}
where $u_{\rm eff}$ is the effective weighted velocity of the plasmon in the first wire, $\omega_c$ is the UV cut-off, and  $\lambda_{11}$ is the element of the matrix $\check{\lambda}=\pi/4e^{2}(\check{C}^{-1}\check{\mathcal{L}})^{-1/2}$. The latter parameter describes the strength of logarithmic QPS–anti-QPS interactions in the first wire, modified by the presence of mutual capacitance $C_{m}$ \cite{LSZ,latyshev2020superconductor}.

The formula (\ref{VVSI}) generalizes the earlier result of Ref.\ \cite{ZGO} to the case of two capacitively coupled nanowires. It has two useful limits:
\begin{equation}
\langle V_{1}(I_{1}) \rangle \;\propto\;
\frac{\Phi_{0}u_{\rm eff} \gamma_1^{2}L}
{\omega_{c}^{2\lambda_{11}}}
\begin{cases}
T^{\,2\lambda_{11}-3}\, I_{1}, & T \gg I_{1}\Phi_{0}, \\[6pt]
I_{1}^{\,2\lambda_{11}-2}, & T \ll I_{1}\Phi_{0}.
\end{cases}
\end{equation}
We recall that the voltage in the second wire is given by $\langle V_{2}(I_{1}) \rangle={\cal D}\langle V_{1}(I_{1}) \rangle$, where the drag coefficient ${\cal D}$ is defined in Eq.~(\ref{drag}).

%===============================================================================
\section{Semiclassical interpretation of the drag effect}

To explain the mass-induced drag effect, we employ the argument presented in Ref.~\cite{LSZ}, adapting it to highlight the influence of the mass term on the dynamics of voltage pulse propagation in both wires, initiated by a QPS event in the first wire.  

We begin by defining the phase field configuration in the basis
\begin{eqnarray}
    \hat{\varphi}(x)| \varphi(x)\rangle = \varphi(x) | \varphi(x)\rangle.  
\end{eqnarray}
Immediately after the occurrence of a QPS, the initial state contains a singular kink:
\begin{multline}\label{shift2}
    \hat{U}_{\xi}(x^{\prime})\hat{\varphi}(x)\hat{U}^{\dagger}_{\xi}(x^{\prime})|\varphi \rangle = \left[\varphi(x) + 2\pi \theta_{\xi}(x-x^{\prime})\right]|\varphi(x) \rangle,
\end{multline}
where $\theta_{\xi}(x-x^{\prime})$ is a smoothed Heaviside step function of width on the order of the coherence length $\xi$, and $\hat{U}_{\xi}(x)$ is the unitary operator 
\begin{eqnarray}\label{U}
    \hat{U}_{\xi}(x) = \exp\left(i \int_{-L/2}^{+L/2} dx^{\prime} \,\partial_{x^{\prime}}\hat{\chi}(x^{\prime}) \theta_{\xi}(x^{\prime}-x)\right),  
\end{eqnarray}
which follows from the commutation relations in the charge–phase representation (\ref{CR}).  
It is convenient to rewrite Eq.~\eqref{shift2} in terms of the initial flux configuration $\hat{\Phi}(x)$ using the duality relation \eqref{duality}:
\begin{eqnarray}\label{init}
    \varphi + 2\pi \theta_{\xi}(x-x^{\prime}) \rightarrow \Phi(x) + 2\pi \delta_{\xi}(x-x^{\prime}),
\end{eqnarray}
where $\delta_{\xi}(x-x^{\prime})$ is a smeared delta function of width $\xi$.

Next, we study the time evolution of the voltage configuration $\hat{V}_{j}(t)$ after the initial state (\ref{init}) is created.  From the Gaussian part of the  Hamiltonian, $\hat{H}_{TL}+\hat{H}_{QPS_{2}}$  [see Eqs.\ (\ref{TL1}) and (\ref{mass})], and the commutation relations (\ref{comm}), the equations of motion are
\begin{eqnarray}\label{charge}	
\partial_{t}\hat{\Phi}_l(x,t)&=&\frac{1}{\Phi_0}\!\sum_j[C^{-1}_{lj}\partial_x^2\hat{\chi}_j(x,t) - m^2_{lj}\hat{\chi}_j(x,t)], \\ 
\partial_{t}\hat{\chi}_l(x,t)&=&\Phi_0\sum_j\mathcal{L}_{lj}^{-1}\hat{\Phi}_j(x,t), \label{charge2}
\end{eqnarray}
and the initial configuration (\ref{init}) implies the condition $\partial_{t}\hat{\chi}_{l}(x,0)=\Phi_{0}\mathcal{L}^{-1}_{l1}\delta_{\xi}(x)$.
The matrix $m_{lj}$ entering Eq.~(\ref{charge}) incorporates the plasmonic mass generated in the second wire once QPS proliferate in the regime $\lambda_2<2$, as discussed in Sec.~II. This mass term is absent in the first wire and appears only in the $(2,2)$ component of $m_{lj}$.

\begin{figure}[h]
\centering
\includegraphics[width=0.48\textwidth]{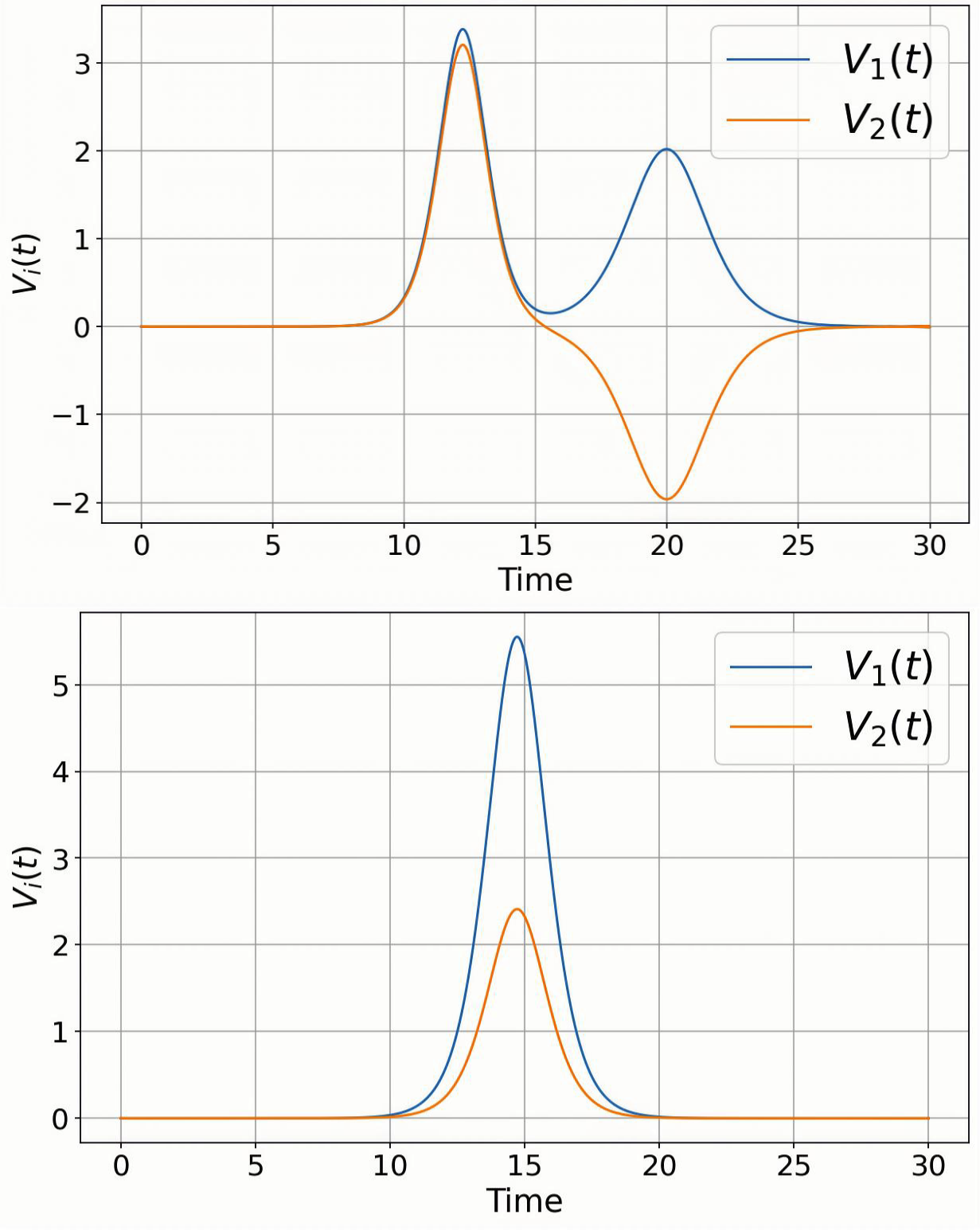}
\caption{Time evolution of the voltage (i.e., the potential drop between two cross-sections at $x=\pm L/2$) in both nanowires, generated by a single QPS in the first wire. Results are shown in arbitrary units for two cases: (a) $m=0$; (b) $m\neq 0$ in the limit $L \gg L_{0}$.}
\label{figure for voltage pulses}
\end{figure}

Solving these equations, the voltage is obtained from Eq.~(\ref{voltage}). Numerical calculations based on Eqs.~\eqref{charge} and \eqref{charge2} are shown in Fig.~\ref{figure for voltage pulses}. Fig.\ \ref{figure for voltage pulses}(a) shows the voltage (\ref{voltage}) as a function of time in both wires for $m=0$. The signal appears when the QPS-generated pulses reach one of the voltage points of the wires at $x=\pm L/2$. As discussed in Ref.~\cite{LSZ}, coupling between the wires splits the plasmon spectrum into a fast charged and a slow dipole mode, with group velocities $v_{\pm}$. Consequently, the voltage splits into two pulses. In the first wire, repeated QPS events lead to a finite time-averaged voltage,
\begin{eqnarray}\label{TA}
  \langle \hat{V}_{1} \rangle=\lim_{\tau\to\infty}\frac{1}{\tau} \int^{\tau/2}_{-\tau/2} dt \; \hat{V}_{1}(t),
\end{eqnarray}
as the two pulses add. In the second wire, the two pulses have opposite sign and cancel exactly, so that $\langle \hat{V}_{2} \rangle=0$ and no drag effect appears. 

Introducing a finite mass term in the Hamiltonian of the second wire dramatically modifies the pulse dynamics, as shown in Fig.~\ref{figure for voltage pulses}(b). Two main effects arise: (1) the velocity difference between the two modes vanishes, causing their pulses to arrive simultaneously and merge into a single peak at distances $L\gg L_0$; (2) the voltage pulse in the second wire remains finite in this limit, yielding a nonzero drag effect $\langle \hat{V}_{2} \rangle\neq 0$.

An alternative interpretation of the drag effect follows from Eq.~(\ref{AV}), where the induced voltage is expressed as a sum of products of two factors: the QPS rate $\Gamma_{QPS}$ and the retarded Green function describing the wire response. In the limit $\omega \to 0$, these factors decouple dynamically. It is well known that for capacitively coupled metals the plasmon-mediated scattering vanishes in this limit, since exciting a plasmon in a finite-size conductor requires finite energy. This explains the absence of drag when both wires remain superconducting. By contrast, once a mass gap opens in the plasmon spectrum of the second wire, the effective coupling between plasmons in the two wires remains finite even as $\omega \to 0$, thereby giving rise to the observed drag effect. This conclusion follows directly from the retarded function in Eq.~(\ref{GF2}).

%===============================================================================
\section{Conclusion}
\label{Conclusion}

We have investigated Coulomb drag in a system of capacitively coupled superconducting nanowires, focusing on the interplay between quantum phase slips (QPS) and collective plasmon modes. Using a combination of perturbative and semiclassical approaches, we demonstrated that when both wires are superconducting the induced voltage in the passive wire vanishes due to exact cancellation of plasmon contributions. In contrast, when the second wire is tuned below the superconductor–insulator transition and acquires a mass gap, this cancellation is lifted and a finite drag voltage emerges. The resulting effect constitutes a mechanism of mass-induced Coulomb drag in low-dimensional superconductors.

Our theory describes the crossover from weak drag in short wires to strong drag in long wires, without assuming weak interactions. The semiclassical analysis provides an intuitive picture: the mass term enforces synchronization of plasmon modes and prevents complete cancellation of voltage pulses, thereby yielding a robust nonlocal response. Together, these results clarify how dimensionality, mutual capacitance, and a plasmonic gap combine to produce finite drag in superconducting nanostructures.

Beyond its conceptual significance, this work identifies new routes for probing correlated dynamics in low-dimensional superconductors. The predicted mass-induced drag could be tested in nanowire devices near the superconductor–insulator transition, offering a direct experimental handle on QPS dynamics and on the role of collective modes in mediating nonlocal transport.

\section*{Acknowledgments}     
We acknowledge the financial support from the Swiss National Science Foundation. 
%=========================================================
\bibliography{refs}

\appendix

\section{The details of the calculations using the Keldysh technique}
\label{Keldysh}

In this Appendix we provide the technical details of the perturbative expansion in $\hat{H}_{QPS_1}$ using the Keldysh formalism, leading to the results summarized in Sec.~\ref{Spectral density of current noise}. The derivation proceeds in three steps: (i) algebra of vertex operators, (ii) evaluation of Green’s functions of the quadratic Hamiltonian, and (iii) calculation of the QPS decay rate and voltage response. 

% Clarified vertex operator algebra
Since $\hat{H}_{QPS_1}$ is a linear combination of vertex operators $\hat{L}_{1}^{\pm}(x,t)=e^{\pm i \hat{\chi}_{1}(x,t)}$, the perturbative expansion relies on their commutation algebra. Using the Baker–Hausdorff formula one obtains
\begin{multline}\label{BH}
 \hat{L}_{j_{1}}^{\sigma_{1}}(x_{1},t_{1}) \hat{L}_{j_{2}}^{\sigma_{2}}(x_{2},t_{2})
 =e^{-\frac{\sigma_{1}\sigma_{2}}{2}[\hat{\chi}_{j_{1}}(x_{1},t_{1}),\hat{\chi}_{j_{2}}(x_{2},t_{2})]} \\
 \times  \hat{L}_{j_{1}j_{2}}^{\sigma_{1}\sigma_{2}}(x_{1},t_{1};x_{2},t_{2}), \qquad \sigma_{k}=\pm 1, 
\end{multline}
where the generalized vertex operator is defined as
\begin{eqnarray}\label{composite}
\hat{L}_{j_{1}\ldots j_{m}}^{\sigma_{1}\ldots
 \sigma_{m}}(x_{1},t_{1},\ldots x_{m},t_{m})=e^{i\sum^{m}_{n=1}\sigma_{n}\hat{\chi}_{j_{n}}(x_{n},t_{n})}. 
\end{eqnarray}
% Rephrased: makes the logical step explicit
Equations (\ref{PT}), (\ref{BH}), and (\ref{composite}) thus provide a systematic framework for the perturbative expansion in terms of vertex operators. 

% Streamlined explanation of averages
Because the averaging in Eq.~(\ref{PT}) is performed with respect to the quadratic Hamiltonian $\hat{H}_{TL}+\hat{H}_{QPS_2}$, the result can be expressed entirely through Gaussian correlators, i.e., retarded, advanced, and Keldysh Green’s functions:
\begin{multline}
 \langle [\hat{\chi}_{i}(x,t),\hat{\chi}_{j}(x,t)]\rangle\\
 = i(G_{ij}^{R}(x,x^{\prime};t_{1},t_{2})-G_{ij}^{A}(x,x^{\prime};t_{1},t_{2})),  
\end{multline}
\begin{multline}
\langle \hat{L}_{j_{1}\ldots j_{m}}^{\sigma_{1}\ldots
 \sigma_{m}}(x_{1},t_{1},\ldots x_{m},t_{m}) \rangle \\
 = e^{-\frac{i}{2}\sum^{m}_{n,l=1}\sigma_{n}\sigma_{l}G_{j_{n},j_{l}}^{K}(x_{n},x_{l};t_{n},t_{l})}.
\end{multline}

% Transition to Green’s functions
The matrix retarded Green function $\check{G}^{R}(x,x^{\prime},\omega)$ follows from the quadratic Hamiltonian $\hat{H}_{TL}+\hat{H}_{QPS_2}$ and satisfies the inhomogeneous equation
\begin{equation}\label{EoM}
(\partial^{2}_{x}+\check{C}\check{\mathcal{L}}\omega^{2}-\check{C} \check{m}^{2}) \check{G}^{R}(x,x^{\prime},\omega)=\Phi_{0}\check{C}\delta(x-x^{\prime}),
\end{equation}
with solution
\begin{multline}\label{GF2}
		\check{G}^{R}(x,x^{\prime},\omega)=\frac{\Phi^{2}_{0}}{2i}e^{i|x-x^{\prime}|\sqrt{\check{C}\check{\mathcal{L}}\omega^{2}-\check{C} \check{m}^{2}}} \\
\times\left(\sqrt{\check{C}\check{\mathcal{L}}\omega^{2}-\check{C} \check{m}^{2}}\right)^{-1}\check{C},
\end{multline}
where 
\begin{eqnarray}
  \check{m}^{2}=\left(\begin{matrix}
   0  & 0 \\
   0  & m^{2}
  \end{matrix}\right)  
\end{eqnarray}
is the mass matrix, and $m^2 = \gamma_2 \Phi_0^2$. The Keldysh Green function is obtained using the fluctuation–dissipation theorem (FDT),
\begin{multline}
\check{G}^{K}(\omega,x,x^{\prime})=\frac{1}{2}\coth\left(\frac{\sqrt{\check{C}\check{\mathcal{L}}\omega^2 - \check{C}\check{m}^{2}}}{2T}\right) \\
 \times \left(\check{G}^R(\omega,x,x^{\prime}) - [\check{G}^R(\omega,x,x^{\prime})]^\dagger \right).
\end{multline}

% Made transition explicit: from expansion to decay rate
Applying the perturbative expansion in $\hat{H}_{QPS_1}$ to leading order, as outlined in Sec.~\ref{Spectral density of current noise}, yields Eq.~(\ref{AV}) of the main text, where the QPS decay rate enters as
\begin{equation}\label{Gammas}
\Gamma_{QPS}(\omega)=\frac{\gamma_1^2}{4}
\int_{-\infty}^{\infty}\!dx\int_{-\infty}^{+\infty}\!dt\;
e^{i \omega t}\,
\big\langle \hat{L}_{1}^{+}(x,t)\,\hat{L}_{1}^{-}(0,0)\big\rangle,
\end{equation}
Using the Gaussian averages above, the decay rate of the current state in the first wire is
\begin{multline}\label{decay}
\Gamma_{QPS}(\omega)=\frac{\gamma^{2}_{1}}{4} \int_{-\infty}^{+\infty} dx \int_{-\infty}^{+\infty} dt \; e^{i\omega t} e^{iG^K_{11}(x,t)-iG^{K}_{11}(0,0)} \\ 
\times e^{\frac{i}{2} (G^R_{11}(x,t) - G^R_{11}(x,-t))}.
\end{multline}
As shown in Ref.~\cite{ZGO}, this expression is equivalent to evaluating the imaginary part of the free energy using the Im$F$ technique \cite{weiss2012quantum}. Ref.~\cite{semenov2017quantum} established the full equivalence of the two approaches, both of which rely on the analytic properties of Green functions in the time domain. Alternatively, the same decay rate can be obtained via the $P(E)$-function formalism \cite{ingold1992single}, widely used in the theory of charge tunneling in metallic junctions embedded in an electromagnetic environment.

% Transition to response function
Finally, we evaluate the response function $G^{R}_{V_j\chi_1}(\omega,x)$ entering Eq.~(\ref{AV}). From the definition of voltage (\ref{voltage}) and Eqs.~(\ref{GF2}), (\ref{RF}) one finds
\begin{multline}\label{DetectorResponse}
    G^R_{V_j\chi_1}(\omega,x)=\frac{1}{\Phi_0} \;\check{P}_j^T\Big[\check{C}^{-1}\big( \partial_{x_{1}}\check{G}^R(\omega,x_1, x)|_{x_1= L/2} \\
- \partial_{x_{2}}\check{G}^R(\omega,x_2,x)|_{x_2= -L/2} \big)\Big]\check{P}_1,
\end{multline}
where we introduced the projectors
\begin{eqnarray}
 \check{P}_{1}=\left[\begin{array}{cc}
     1   \\
     0  
 \end{array}\right], \;\; \check{P}_{2}=\left[\begin{array}{cc}
     0   \\
     1  
 \end{array}\right].  
\end{eqnarray}
In the zero-frequency limit, this reduces to
\begin{multline}\label{matrixexp}
    \lim_{\omega\to 0}G^R_{V_j\chi_1}(\omega,x)= 
\Phi_0 \check{P}_{j}^{T}\Big[\check{C}^{-1}e^{-L/2\sqrt{\check{C}\check{m}^{2}}} \\
\times \cosh\Big(x\sqrt{\check{C}\check{m}^{2}}\Big)\check{C}\Big]\check{P}_1.
\end{multline}
Substituting into Eq.~(\ref{AV}) yields
\begin{multline}\label{voltage1}
 \langle \hat{V}_{j}(I_{1})\rangle = \Phi_{0}\check{P}^{T}_{j} \\
 \times 
 \left(\begin{matrix}
 L & 0 \\
 \frac{C_{m}L}{C_{2}}-\frac{C_{m}(1-e^{-L\sqrt{C_{2}m^{2}}})}{\sqrt{C^{3}_{2}m^{2}}} & \frac{(1-e^{-L\sqrt{C_{2}m^{2}}})}{\sqrt{C_{2}m^{2}}}
 \end{matrix}\right) \\
 \times \check{P}_{1}\left[\Gamma_{QPS}(I_{1}\Phi_{0})-\Gamma_{QPS}(-I_{1}\Phi_{0})\right]. 
\end{multline}
Evaluating the projectors explicitly reproduces the results of the main text for the voltage in the first wire, Eq.~(\ref{voltage drop}), and the drag coefficient, Eq.~(\ref{drag}).

\end{document}